\documentclass[oldversion]{aa} 
\usepackage{graphicx}
\usepackage{aalongtable}
\usepackage{txfonts}
\usepackage{rotating}
\usepackage{lscape}
\newcommand{\gsim}{\;\lower.6ex\hbox{$\sim$}\kern-7.75pt\raise.65ex\hbox{$>$}\;}
\newcommand{\lsim}{\;\lower.6ex\hbox{$\sim$}\kern-7.75pt\raise.65ex\hbox{$<$}\;}

\begin{document}
\title{An aluminium tool for multiple stellar generations in the globular
clusters 47~Tuc and M~4\thanks{Based on observations collected at ESO telescopes
under programme 085.D-0205 and on public data from the ESO/ST-ECF  Science Archive
Facility.}\fnmsep\thanks{Tables 2 and 3 are only available in electronic form at
the CDS via anonymous ftp to {\tt cdsarc.u-strasbg.fr} (130.79.128.5) or via
{\tt http://cdsweb.u-strasbg.fr/cgi-bin/qcat?J/A+A/???/???}}
 }

\author{
Eugenio Carretta\inst{1},
Raffaele G. Gratton\inst{2},
Angela Bragaglia\inst{1},
Valentina D'Orazi\inst{3,4},
\and
Sara Lucatello\inst{2}
}

\authorrunning{E. Carretta et al.}
\titlerunning{Aluminium abundances in 47~Tuc \& M~4}

\offprints{E. Carretta, eugenio.carretta@oabo.inaf.it}

\institute{
INAF-Osservatorio Astronomico di Bologna, Via Ranzani 1, I-40127
 Bologna, Italy
\and
INAF-Osservatorio Astronomico di Padova, Vicolo dell'Osservatorio 5, I-35122
 Padova, Italy
\and
Department of Physics and Astronomy, Macquarie University, Balaclava rd, North
Ryde, NSW 2109, Australia
\and
Monash Centre for Astrophysics, School of Mathematical Sciences, Building 28,
Monash University, VIC 3800, Australia}

\date{}

\abstract{We present aluminium abundances for a sample of about 100 red giant
stars in each of the Galactic globular clusters 47~Tuc (NGC~104) and M~4
(NGC~6121). We have derived homogeneous abundances from intermediate-resolution
FLAMES/GIRAFFE spectra. Aluminium abundances are from
the strong doublet Al {\sc i} 8772-8773~\AA\ as in previous works done for
giants in NGC~6752 and NGC~1851, and nitrogen abundances are extracted from a
large number of features of the CN molecules, by assuming a suitable carbon
abundance. We added previous homogeneous abundances of O and Na and newly derived
abundances of Mg and Si for our samples of 83 stars in M~4 and 116 stars in
47~Tuc to obtain the full set of elements from proton-capture reactions
produced by different stellar generations in these clusters. By
simultaneously studying the Ne-Na and Mg-Al cycles of H-burning at high
temperature our main aims are to understand the nature of the polluters at
work in the first generation and to ascertain whether the second generation
of cluster stars was formed in one or, rather, several episodes of star
formation. Our data confirm that in M~4 only two stellar populations are
visible. On the other hand, for 47~Tuc a
cluster analysis performed on our full dataset suggests that at least three
distinct groups of stars are present on the giant branch. 
The abundances of O, Na, Mg and Al in the intermediate group can be
produced within a pollution scenario; results for N are ambiguous, depending
on the C abundance we adopt for the three groups.
}

\maketitle

\section{Introduction}

The new paradigm for globular cluster (GCs) formation is currently well
assessed in its general aspects. Several decades of spectroscopic observations
as well as years of high precision imaging mainly with $HST$\ have
shown that a typical pattern is present within each galactic GC investigated
so far (see the recent reviews by Gratton, Sneden and Carretta 2004; Martell
2011; Gratton, Carretta and Bragaglia 2012a and references therein). A first generation of stars
is found side by side with another, slightly younger stellar generation
whose components are unambiguously formed by ejecta processed in the
nuclear H-burning at high temperature within massive stars of the first
generation in each cluster (see e.g. Gratton et al. 2001 for the first
clear-cut observational constraint - followed by Ramirez and Cohen 2002,
Carretta et al. 2004 -, and Ventura et al 2001 and Decressin et al
2007 for different views on the nature of these early polluters of the
intra-cluster medium).

The concept of GCs being good examples of Simple Stellar Populations (SSP)
has been eventually replaced by their being made of
Multiple Stellar Populations (MSP) whose compositions are marked by a typical
pattern: light elements show anti-correlations and correlations
among species produced (such as Na, Al, Si) and destroyed (such as O,
Mg, F) in the network of proton-capture reactions responsible for the
nuclear processing of the polluting matter (e.g. Denisenkov \& Denisenkova
1989; Langer et al. 1993, Arnould et al. 1999). In almost all GCs studied
so far, ranging from the most massive one to the smallest objects in the
Harris catalogue (1996, and following updates), this pattern is retrieved and it is
so typical a signature of GCs that a new recent definition of {\it bona fide}
globular has been proposed entirely based on the presence of the most notable signature,
the Na-O anti-correlation (Carretta et al. 2010), discovered by the Lick-Texas
group in the 1990s (see the review by Kraft 1994) and extensively studied in many
clusters (see the updated review by Gratton et al. 2012a and references
therein.).

While the first level of the game is clear, several details still are left
in the poorly known age when GCs formed about a Hubble time ago. Existent data are
not enough to uncover yet the true nature of the polluters that provided the
material for the building up of the second generations. The most favourite
candidates could be both intermediate-mass AGB stars (Ventura et al. 2001) or
very massive, fast rotating stars (Decressin et al. 2007), although alternative
ones have been presented, such as massive binaries (de Mink et al. 2009).
Moreover, the distributions of stars along e.g. the Na-O anti-correlations
do require that the matter composing the second stellar generation must be
diluted in some amount with material that did not experience hot nuclear burning.
This rather strong constraint stems from both observations (the relations
between proton-capture elements such as O and Na and the fragile element Li,
easily destroyed at temperatures much lower than those involved in the
hot H-burning, see Prantzos and Charbonnel 2006 and the discussion in
Gratton et al. 2012a) and from theory (the yields
from AGBs would result necessarily into a correlation, unless some amount
of dilution is involved, D'Ercole et al. 2011).

Leaving aside for the moment several problems related to the nature of the
diluting matter (D'Ercole et al. 2011), it is important to
ascertain whether a model of dilution does really work and what are the observational
consequences concerning the formation of multiple stellar generations in GCs.
Two scenarios can be envisioned:
\begin{enumerate}
\item After the formation of the first generation, a single mixture of matter
from polluters diluted with unprocessed matter produces a second generation
of stars distributed along the Na-O and Al-Mg anti-correlations. Since what
is changing along these distributions is the degree of dilution, we expect
that elements destroyed in the proton-capture reactions (namely O and Mg) or
species enhanced in these processes (i.e. Na and Al) show a direct, linear
proportionality.
\item On the other hand, several episodes of star formation might have occurred 
in the proto-GC after the formation of the first generation. Each one of them 
might have been characterized by a typical polluter with a given average 
mass. In this case, we expect that the elements are not necessarily produced or
destroyed in linear proportion with each other. The distributions might still 
grossly resemble an anti-correlation, but different groups might be distinguished.
\end{enumerate}

Ideally, we would like to have for these studies a database composed of many
stars  with detailed abundances of several light elements. This requires
extensive sets  of very high resolution and S/N spectra with large spectral
coverage, such as the  FLAMES/UVES spectra obtained by Marino et al. (2008) who
provided abundances for a good  number of proton-capture elements (O, Na, Mg,
and Al for 93, 105, 105, and 87 stars,  respectively) for M~4. When compared to
our analysis for this cluster based on moderate  resolution GIRAFFE spectra
(Carretta et al. 2009a), their analysis show a clumpy  distribution that was
masked by noise on our lower quality data, despite the small  star-to-star
errors derived by our tailored procedure. Unluckily, it is not easy to  obtain
such high quality material for a large number of clusters because such 
observations are very expensive in telescope time. 

On the other hand, much insight can still be obtained by extending the spectral
range  of our moderate resolution survey to other spectral ranges, allowing to
observe  transitions for elements whose data were missing in our original
spectra. This can be  obtained with a very limited additional observing time
exploiting the GIRAFFE spectrograph, which allows simultaneous observation of
more than a hundred star in a cluster, and can then be done easily for several
clusters. Using this technique, we  showed e.g. that the globular cluster
NGC~6752 is made of three discrete populations  (Carretta et al. 2012a; 
note that the presence of three groups of stars in NGC~6752 is quite clear
e.g. in the results obtained by Yong et al. 2008, but they did not make any
comment about this in their paper).
Moreover, we could show that the composition of the intermediate  population of
that cluster cannot be reproduced by simply mixing suitable amounts of the 
material which made the two other populations. This clearly points towards the
second scenario mentioned above, possibly supporting the finding by Yong et al. (2008)
that contributions from both AGB and massive stars are required to explain the
correlations in NGC~6752.

The present paper presents similar results on Al for two additional GCs
(47~Tuc=NGC~104  and M~4=NGC~6121) for which we have already produced Na, O
abundances in our FLAMES  survey. The analysis is similar to that presented in
Carretta et al. (2012a) for  NGC~6752 and Carretta et al. (2012b) for
NGC~1851.

While for M~4 our data largely overlap with those of Marino et al. (2008), for 
47~Tuc, this is the largest sample of RGB stars with homogeneous  Na, O and Al
abundances that we are aware of; the previous largest sample observed at
medium-high resolution is represented by
the 14 stars  with UVES spectra analysed by our group in Carretta et al.
(2009b), for Al, and Carretta et al. (2009a) concerning Na and O abundances.

\section{Observations and analysis}

The present work is based on two exposures, the first of 2700 sec at airmass
1.487 (seeing=1.23") acquired on 2010, July 30 for 47~Tuc, and the second of 1600
sec at airmass 1.240 (seeing=1.79") taken on 2010, April 14 with FLAMES@VLT-UT2
and the high resolution grating HR21. The resolution is 17,300 at the center of
the spectra and the spectral range goes from about 8484 to about 9001~\AA,
including the  Al~{\sc i} doublet at 8772-73~\AA, which is the main feature we
were interested in. The positioning of fibers used the same
configurations employed to observe the strong doublet of Na (at 5682 and 5688~\AA)
with the grating HR11 in Carretta et al. (2009a) to maximize the number of stars
in the two clusters along the expected Na-Al correlation. We observed 116 giants
in 47~Tuc in the magnitude range $V=12.25-14.57$ and 83 RGB stars in the
magnitude range $V=11.63-14.08$ in M~4. Coordinates and magnitudes for all stars
in our  samples can be retrieved in the on-line tables of Carretta et al.
(2009a).

Data reduction was performed by the ESO personnel through the dedicated ESO
FLAMES-GIRAFFE pipeline. The output bias-corrected, flat-fielded,
one-dimensional and wavelength-calibrated spectra were sky subtracted and
shifted to zero radial velocity using IRAF\footnote{IRAF is the Image Reduction
and Analysis Facility, a general purpose software system for the reduction and
analysis of astronomical data. IRAF is written and supported by the IRAF
programming group at the National Optical Astronomy Observatories (NOAO) in
Tucson, Arizona. NOAO is operated by the Association of Universities for
Research in Astronomy (AURA), Inc. under cooperative agreement with the National
Science Foundation.}.

The signal-to-noise (S/N) ratio of individual spectra is listed in
Table~2 and Table~3 (only available on-line) for 47~Tuc
and M~4, respectively. The median S/N is 249 for spectra of stars in 47~Tuc and
168 for giants observed in M~4.

\section{Analysis and derived abundances}

The atmospheric parameters required for the analysis were simply adopted from
Carretta et al. (2009a) where a full abundance analysis was done for all the
stars targeted in the present study.
This approach has two main advantages: first, it provides a full set of
homogeneous parameters, with very small star-to-star errors\footnote{
Typical internal errors are 4-6 K in temperature, 0.04 dex in gravity, 0.03 dex
in [Fe/H] and 0.02-0.03 km/sec in microturbulent velocity, see Table A.2 in
Carretta et al. 2009a.}, useful to minimise
down a possible smearing along the observed anti-correlations due only to errors
associated to the analysis. Moreover, the number of lines of iron with well known
atomic parameters is not high in the spectral range of the grating HR21, whereas
the [Fe/H]\footnote{We adopt the usual spectroscopic notation, $i.e.$  for any
given species X, [X]=  $\log{\epsilon(X)_{\rm star}} - \log{\epsilon(X)_\odot}$
and   $\log{\epsilon(X)}=\log{(N_{\rm X}/N_{\rm H})}+12.0$\ for absolute number
density abundances.} values derived in Carretta et al. (2009a) rest on several
iron lines. Second, O and Na abundances, homogeneously derived using the same
atmospheric parameters, can be safely coupled with the new abundances of Al, Mg,
and Si.

The complete list of atmospheric parameters (effective temperature T$_{\rm
eff}$, surface gravity, metallicity and microturbulent velocity), as well as the
typical associated errors,  can be retrieved from the on-line tables in Carretta
et al. (2009a) for all stars in 47~Tuc and M~4 analysed in the present study.

\subsection{Nitrogen}

CN lines are ubiquitous in the observed spectral range,
and some of them could also contaminate the Al lines. Thus, the first
step is to reproduce the CN features as much accurately as possible. The problem
is more severe in metal-rich clusters such as 47~Tuc ([Fe/H]$=-0.77$ dex,
Carretta et al. 2009c), but some degree of contamination should also be
expected in the more metal-poor cluster M~4 ([Fe/H]$=-1.17$ dex, Carretta et
al. 2009c).

To deal with this aspect, we used an improved version of the procedure adopted 
in Carretta et al. (2012b). We first obtained a coadded spectrum with very high 
S/N value by summing up the ten spectra of 47 Tuc stars with the highest S/N. 
On this master spectrum, we individuated eight regions apparently free of lines
to be used to derive a local reference continuum and 18 spectral regions
dominated by CN bands. Definitions of these spectral regions and of those
for the local reference continuum are given in Tab.~\ref{t:feature}.

\begin{table}
\centering
\caption[]{List of regions used to estimate local continuum and fluxes of CN
bands}
\begin{tabular}{lccc}
\hline
      & $\lambda_{in}$  & $\lambda_{fin}$  & $\Delta \lambda$  \\
      &    \AA          &      \AA         &     \AA           \\
\hline        
  CN  &	8759.22 & 8760.22 & 1.00  \\
  CN  &	8762.66 & 8763.40 & 0.74  \\
  CN  &	8767.24 & 8768.24 & 1.00  \\
  CN  &	8769.11 & 8769.92 & 0.81  \\
  CN  &	8769.96 & 8771.34 & 1.38  \\
  CN  &	8774.84 & 8776.12 & 1.28  \\
  CN  &	8782.88 & 8783.82 & 0.94  \\
  CN  &	8788.92 & 8789.61 & 0.69  \\
  CN  &	8794.02 & 8794.76 & 0.74  \\
  CN  &	8799.14 & 8800.06 & 0.92  \\
  CN  &	8801.77 & 8802.65 & 0.88  \\
  CN  &	8812.42 & 8813.42 & 1.00  \\
  CN  &	8813.42 & 8814.01 & 0.59  \\
  CN  & 8815.63 & 8816.52 & 0.89  \\
  CN  & 8819.04 & 8820.32 & 1.28  \\
  CN  & 8822.12 & 8823.38 & 1.26  \\
  CN  & 8830.32 & 8831.44 & 1.12  \\
  CN  & 8835.07 & 8836.86 & 1.79  \\
cont  & 8739.29 & 8739.71 & 0.42  \\
cont  & 8746.22 & 8746.53 & 0.31  \\
cont  & 8765.41 & 8765.60 & 0.19  \\
cont  & 8771.32 & 8771.65 & 0.33  \\
cont  & 8791.26 & 8791.46 & 0.20  \\
cont  & 8791.74 & 8792.12 & 0.38  \\
cont  & 8794.76 & 8795.32 & 0.56  \\
cont  & 8834.69 & 8835.14 & 0.45  \\  
\hline
\end{tabular}
\label{t:feature}
\end{table}

In the second step, synthetic spectra were computed using the Kurucz (1993) grid
of model atmospheres (with the overshooting option switched on) and the same line list
used in Carretta et al. (2012b), using the atmospheric parameters suitable for
each individual star in 47~Tuc. Three spectra were computed assuming
[C/Fe]=-0.6 dex. and three evenly spaced values of nitrogen, [N/Fe]=-0.5,
+0.25 and +1.0 dex, with the package ROSA (Gratton 1988).
To each of the 18 CN features in each star we applied the procedure explained in
Carretta et al. (2012b): the average flux within the in-line region of the
feature was measured and a weighted reference continuum was derived from all the
8 off-line regions using weights equal to the width of each region. Whenever a defect in
the spectrum was found on a continuum region from visual inspection, that region
was excluded from the computation of the reference local continuum.
The N abundance from each feature was then obtained by comparing the normalized
flux with the fluxes measured in the same way on the three synthetic spectra.

The above procedure was repeated for each of the 116 stars observed in 47~Tuc
and for each star an average [N/Fe] value was derived.
The spectral regions of each CN feature were inspected visually and the relative
feature discarded from the average whenever a spike was found to fall on the
feature.
The final abundances of N were obtained after applying a k$\sigma$-clipping at
2.5 $\sigma$ to the average abundance from individual features in each star.
These final abundances in 47~Tuc rest on average on 17 features. We found a
mean value [N/Fe]$=+1.038 \pm 0.011$ dex (rms=0.119 dex, 116 stars).

This approach was adopted also for the 83 stars observed in M~4; also in this
case, the mean N abundance was derived from an average of 17 CN features. 
finding [N/Fe]$=+1.052 \pm 0.017$ dex (rms=0.152 dex, 83 stars). 
The larger scatter observed in M~4 is mainly due to the giants in
this more metal-poor cluster being on average 200 K warmer than giants in
47~Tuc (and only minimally due to the lower S/N of the spectra). The molecular
features are more difficult to measure and the warmest stars are thus
responsible for most of the observed scatter. 

Individual values of N abundances are listed in Tab.2 and
Tab.3 for 47~Tuc and M~4, respectively. We recall that
these [N/Fe] were obtained assuming [C/Fe]=-0.6. Were the C abundance for
a given star be different from this value, [N/Fe] should be modified
in such a way that the sum [C/Fe]+[N/Fe] remains constant, because the 
concentration of CN depends on the product of the C and N concentrations.

The C abundances are expected to change as a function of temperature due to 
deep mixing. We estimated from C and N abundances measured in red giants of 
M~4 by Ivans et al. (1999) that over the temperature range of our sample the 
maximum variation in C expected as a function of the luminosity is about 
0.25~dex. The effect of star-to-star variations in C abundances at a given
luminosity, related to the multiple population phenomenon, will be discussed later.
The impact is likely very important in the case of 47 Tuc. For M~4, it seems less
of a problem. For instance, we have sixteen stars in common with Smith \& Briley 
(2005) that obtained C abundances from low resolution spectra. If we divide these 
stars in 6 Na-poor and 10 Na-rich stars (at [Na/Fe]=0.4), 
we find [C/Fe]=$-0.57\pm 0.04$\ (r.m.s.=0.10) and [C/Fe]=$-0.64\pm 0.06$~dex 
(r.m.s=0.20~dex) for the two groups, both compatible with our assumption of [C/Fe]=-0.6.

\begin{figure}
\centering
\includegraphics[bb=20 162 577 448, clip, scale=0.45]{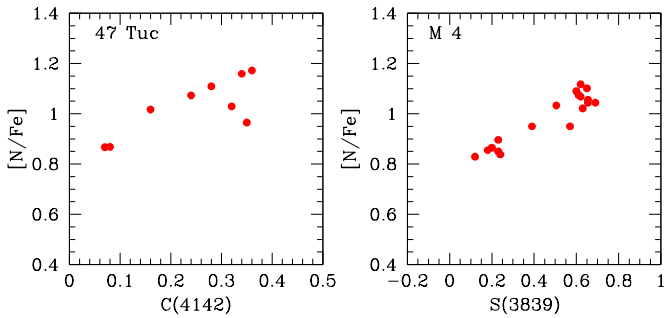}
\caption{Left: comparison of our derived [N/Fe] ratios with the values of the
DDO index from Norris and Freeman (1979) in 47~Tuc. Right: comparison of our N
abundances with the S(3839) index for the strength of the $\lambda$3883 CN band
from the compilation by Smith and Briley (2005) in M~4.}
\label{f:cfrCN}
\end{figure}

We tested the adopted procedure on the spectrum of Arcturus\footnote{We 
retrieved this spectrum in the relevant wavelength range from the ESO Advanced 
Data Product Archive ($http://archive.eso.org/eso/eso\_archive\_adp.html$).}. 
We obtained [N/Fe] = 0.68
adopting [C/Fe]=0 and the atmospheric parameters from Hinkle et al. (2000) and
Carlberg et al.  (2012): T$_{\rm eff}=4300$~K, $\log{g}=1.70$, [Fe/H]=-0.55, and
$v_t=1.64$). This  result is very similar to that obtained by D'Orazi et al.
(2011) with very similar atmospheric parameters.

For a check, we searched the literature for analysis with published  values of N
or CN (or related indexes) in these two clusters. We found 11 RGB stars of
47~Tuc in common with the sample studied by Norris and Freeman (1979), and we
measured N abundances for 9 of them; we also have 29 stars in common with the
compilation of CN values by Smith and Briley (2005) in M~4, with our N values
available for 18 of them. In Fig.~\ref{f:cfrCN} we show the comparison of our
derived N abundances with the values of the DDO index C(4142) in 47~Tuc (left
panel) and of the re-homogenized index S(3839) by Smith and Briley (2005) in
M~4 (right panel). In both cases, with our assumptions about the carbon
abundance, our values are quite well correlated with these literature values. We
conclude that our values provide a good estimate of the N content in the program
stars.

\subsection{Aluminium, magnesium and silicon}

A similar procedure was used to derive Al abundances (see also Carretta et
al. 2012b). We used for each star three synthetic spectra computed with the
appropriate atmospheric parameters (from Carretta et al. 2009a) and the N
abundances derived as described above. The fluxes for determining Al were
measured in the in-line region from 8772.4 to 8774.7~\AA\, while two slightly
different regions were used in the two clusters to derive the local reference
continuum owing to the different metallicity of these GCs. The selected regions
were 8771.28-8771.70~\AA\ and 8776.54-8777.46~\AA\ in 47~Tuc, and
8768.90-8770.00~\AA\ and 8775.20-8776.10~\AA\ in M~4.
Again, we obtained a final continuum by averaging the contribution from the
two regions with weights given by the number of pixels (i.e. the widths of these
intervals).

Abundances of Al were then derived by interpolating the normalized fluxes in the
in-line Al region among those obtained from the synthetic spectra, computed for
[Al/Fe]=0.0, 0.75, and 1.5 dex. As above, should one of the reference local
continuum regions or one of the Al lines be affected by a defect, following visual
inspection, the contaminated region would be dropped from the computation.

Abundances of Mg and Si were obtained from the same set of lines described in
Carretta et al. (2012a), using atmospheric parameters derived in Carretta et al.
(2009a).

Final Al, Mg, and Si abundances for giants in 47~Tuc and M~4 are listed in
Tab.2 and Tab.3, respectively.

\subsection{Errors}

Star-to-star errors in the adopted atmospheric parameters were estimated in
Carretta et al. (2009a) and are shown in Tab.~\ref{t:sensitivity}. To
translate these errors into errors on the Al abundances, we need the sensitivities of Al to
changes in the adopted parameters. These were obtained by repeating the
analysis for the stars 29490 (T$_{\rm eff}=4584$ K) in 47~Tuc and star 30450
(T$_{\rm eff}=4622$ K) in M~4\footnote{These stars were chosen because they have
an effective temperature in the middle of the temperature range of their
respective samples.} and changing a parameter each time, holding the others
fixed at their original value. The amount of the variations and the sensitivity
of each parameter to these changes are listed in Tab.~\ref{t:sensitivity}.

\begin{table*}
\setcounter{table}{3}
\centering
\caption[]{Sensitivities of Al to variations in the atmospheric
parameters and to errors in fluxes, and errors in
abundances [A/Fe] for stars in 47 Tuc and M 4}
\begin{tabular}{lrrrrrrr}
47 Tuc      &         &               &          &         &          &         &            \\
\hline
Element     & [N/Fe]  & T$_{\rm eff}$ & $\log g$ & [A/H]   & $v_t$    & flux     & Total     \\
            &         &      (K)      &  (dex)   & (dex)   &kms$^{-1}$& (dex)   &Internal    \\
\hline        
Variation&   +0.200    &  50           &   0.200   &  0.100   &  0.10    &         &         \\
Internal &   +0.012    &   6           &   0.020   &  0.032   &  0.11    & 0.013   &         \\
\hline
$[$Al/Fe$]${\sc i}&  0.077 & $-$0.032      &   0.000   & 0.115 &        +0.013 &   &0.042    \\
\hline
            &         &               &          &         &          &         &            \\
            &         &               &          &         &          &         &            \\
M 4         &         &               &          &         &          &         &            \\
\hline
Element     & [N/Fe]  & T$_{\rm eff}$ & $\log g$ & [A/H]   & $v_t$    & flux     & Total     \\
            &         &      (K)      &  (dex)   & (dex)   &kms$^{-1}$& (dex)   &Internal    \\
\hline        
Variation&   +0.200    &  50           &   0.200   &  0.100   &  0.10    &         &         \\
Internal &   +0.028    &   4           &   0.041   &  0.025   &  0.12    & 0.019   &         \\
\hline
$[$Al/Fe$]${\sc i}&  0.013 & $-$0.031      &   0.030   & 0.104 &        +0.012 &   &0.036    \\
\hline

\end{tabular}
\label{t:sensitivity}
\end{table*}

The star-to-star (internal) errors in the Al abundances are obtained by summing
in quadrature all the contributions, including the impact of errors in the
derivation of N abundances from CN features.

We attached an error due to flux measurement to the Al abundances. This was
evaluated, as in Carretta et al. (2012b) estimating photometric errors from the
S/N ratio of the spectra of stars and from the width within each of the
reference continuum and in-line regions (and then the number of independent pixels
used). The error in Al abundances is then obtained by comparing these Al
abundances with those derived considering a new value of the Al line strength
index that is the sum of the original values and its error. The errors from
these procedure are listed for each star in Tab.2 and
Tab.3; on average, they are $+0.013 \pm 0.001$ dex from 116 stars
in 47~Tuc and $+0.019 \pm 0.001$ dex from 83 giants in M~4.

Moreover, we applied the same approach to one of the CN features used to
estimate the N abundance. The average error is 0.012 dex in 47~Tuc and +0.028
dex in M~4. These values were assumed as conservative estimates in the error
budget, since final N abundances were obtained from a large number of features.

Summing up all the above contributions the typical internal errors in the
derived Al abundances are 0.042 dex and 0.036 dex for 47~Tuc and M~4,
respectively.
Typical internal errors in [Mg/Fe] and [Si/Fe], estimated according to
the usual procedure (Carretta et al. 2009a), are 0.10 dex and 0.06 dex,
respectively, for both clusters. 

\section{Results}
\begin{figure}
\centering
\includegraphics[scale=0.50]{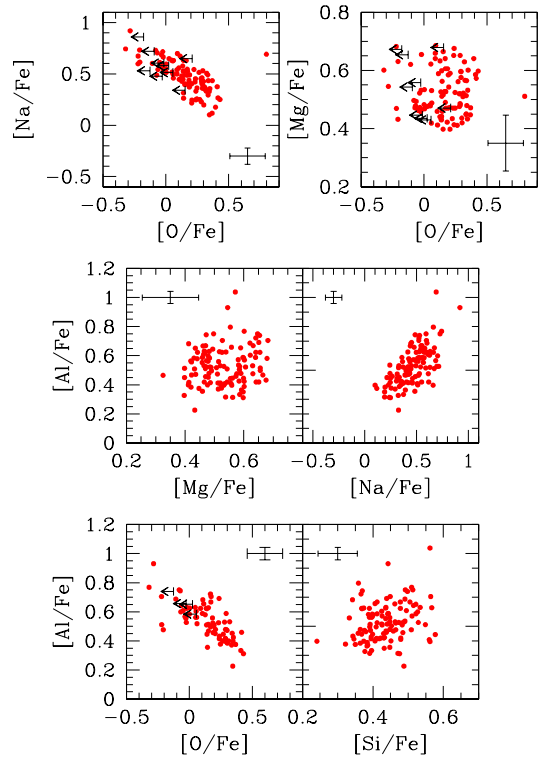}
\caption{Run of the abundance ratios of proton-capture elements O, Na, Mg, Al
and Si among red giants in 47~Tuc, from Carretta et al. (2009a) and the present
study. Upper limits in O are indicated by arrows. Typical star-to-star errors
are also indicated.}
\label{f:fig10a}
\end{figure}

\begin{figure}
\centering
\includegraphics[scale=0.50]{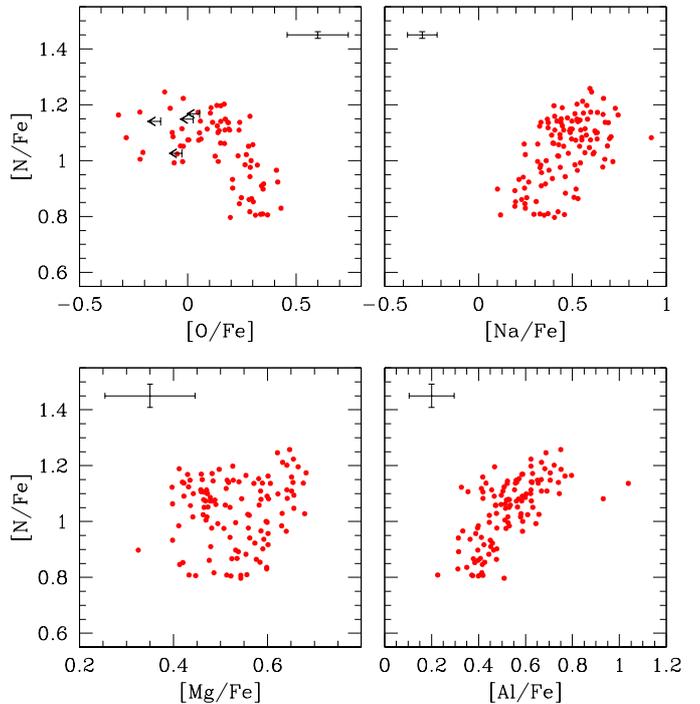}
\caption{Run of the nitrogen abundances as a function of O, Na, Mg, and Al for
our sample of RGB stars in 47~Tuc.}
\label{f:figN10}
\end{figure}

The pattern of abundances of proton-capture elements among red giants in 47~Tuc
is summarized in Fig.~\ref{f:fig10a} and Fig.~\ref{f:figN10}.

The Na-O anti-correlation in 47~Tuc was already known to be rather short, when
considering its large total mass, which is one of the main parameters driving
the extent of the anti-correlation (Carretta et al. 2010). The new set of light
elements added in the present work confirms this trend: Mg and Si do not show
large star-to-star variations, the average values being [Mg/Fe]$=+0.532 \pm
0.007$ dex (rms=0.079, 147 stars) and [Si/Fe]$=+0.440 \pm 0.005$ dex (rms=0.065,
147 stars)\footnote{The number of stars where Mg and Si is measured is higher
than those with Al abundances since transitions of Mg and Si are available in
the spectral range of both the gratings HR11 and HR13, whereas Al from GIRAFFE
spectra is available for only the subsample observed with the grating HR21.}.
The aluminium abundances span a range of about 0.5 dex among giants in this
cluster, with an average value of [Al/Fe]$=+0.529 \pm 0.012$ dex (rms=0.132, 116
stars). Clear trends of Al ratios correlated to Na and anti-correlated to O
abundances show the presence in second generation stars of ejecta processed
through the high temperature Mg-Al cycle. This finding is also confirmed by the
correlation (shallow, but still statistically robust) between Si and Al, that by
itself points toward temperature of H-burning in excess of $\sim 65 \times 10^6$
K, where the leakage from the Mg-Al cycle on $^{28}$Si is onset (see Yong et al.
2005, Carretta et al. 2009b).

As expected, the abundances of N derived from the transformation of O in the CNO
cycle, are nicely correlated with elements enhanced in proton-capture reactions
(such as Na, Al) and anti-correlated with oxygen
(Fig.~\ref{f:figN10}). The apparent lack of a N-Mg anti-correlation is not
a big source of concern, since the Mg variations are small. Our survey of
Al and Mg abundances in several GCs, based on limited samples of RGB stars
observed with FLAMES-UVES (Carretta et al. 2009b) show that Al-rich and
Mg-depleted stars are present only in massive (NGC~2808, NGC~6388, NGC~6441) or
metal-poor (NGC~6752) clusters, or both (NGC~7078=M~15). In such clusters 
a clear Mg-Al anticorrelation is observed even among main sequence stars
(see e.g. Bragaglia et al. 2010 for NGC~2808). In GCs such as those under
scrutiny, small and/or metal-rich, large star-to-star variations of Mg
abundances are not expected, as shown by Carretta et al. (2009b) in a
survey of limited samples of giants in 15 GCs and by Marino et al. (2008) over
about 100 giants in M~4. 

\begin{figure}
\centering
\includegraphics[scale=0.50]{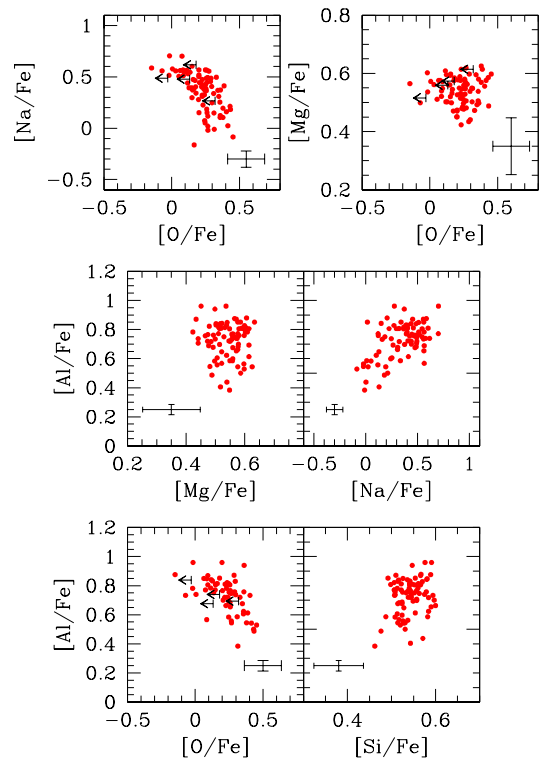}
\caption{The same as in Fig.~\ref{f:fig10a} for M~4.}
\label{f:fig61a}
\end{figure}

\begin{figure}
\centering
\includegraphics[scale=0.50]{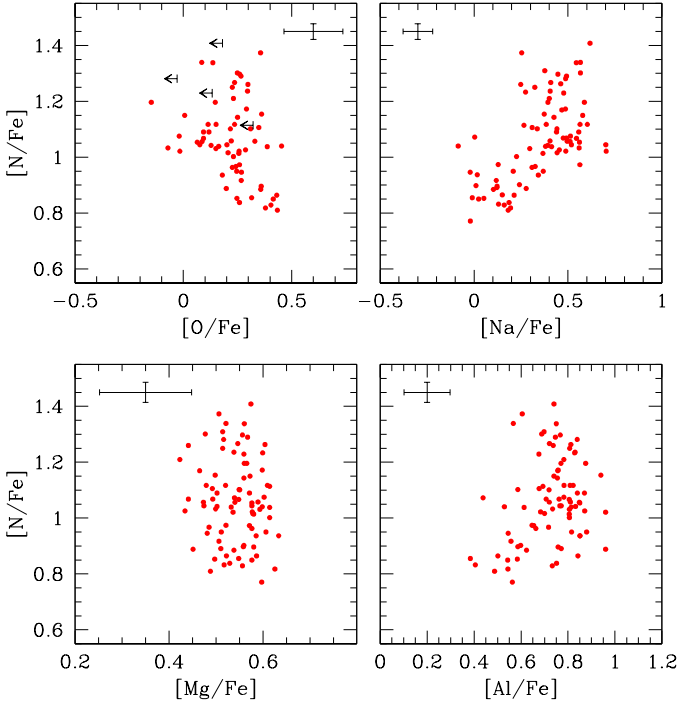}
\caption{The same as in Fig.~\ref{f:figN10} for M~4.}
\label{f:figN61}
\end{figure}

The analogous run of light elements in the globular cluster M~4 are shown in
Fig.~\ref{f:fig61a} and Fig.~\ref{f:figN61}. In this small mass cluster the Na-O
anti-correlation is found to be of moderate extension (Carretta et al. 2009a,
Marino et al. 2008, Ivans et al. 1999) and
also star-to-star variations in Mg and Si are not large. The average values from
our analysis are [Mg/Fe]$=+0.541 \pm 0.005$ dex (rms=0.048, 103 stars) and
[Si/Fe]$=+0.540 \pm 0.003$ dex (rms=0.034, 103 stars). As already indicated by
previous analysis (Marino et al. 2008, Carretta et al. 2009b), star-to-star
variation in Al abundances are not large along the RGB in M~4. A clear Na-Al
correlation is evident in our data, and in the UVES data of the large
sample in Marino et al. (2008), whereas it cannot be seen in the limited sample
of stars analyzed in Carretta et al. 2009b). On the other hand, no correlation 
between Si and Al can be appreciated in the present study.

\begin{figure}
\centering
\includegraphics[bb=19 399 477 681, clip, scale=0.52]{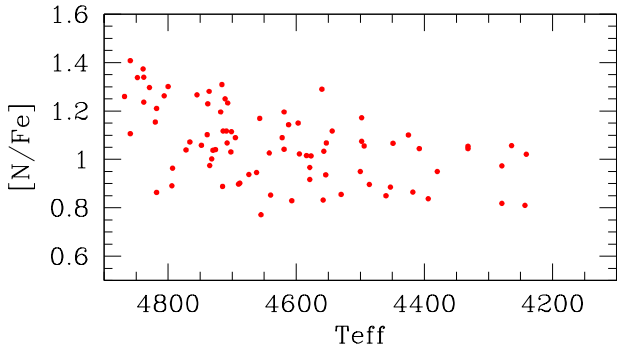}
\caption{Abundances of N as a function of effective temperature in giants of M~4}
\label{f:cnteff61}
\end{figure}

Concerning N abundances, we found a somewhat larger scatter with respect to the
results of 47~Tuc. As explained above, this is largely due to the
increased difficulty to measure CN features in the warmest stars of our sample
in this more metal poor cluster. This is shown in Fig.~\ref{f:cnteff61} where
[N/Fe] ratios are plotted as a function of the effective temperature: the
scatter noticeably increases for stars at the warmest end of the sample, which are
those accounting for most of the dispersion seen in Fig.~\ref{f:figN61}.

\subsection{Comparison with previous studies}

\begin{figure}
\centering
\includegraphics[scale=0.50]{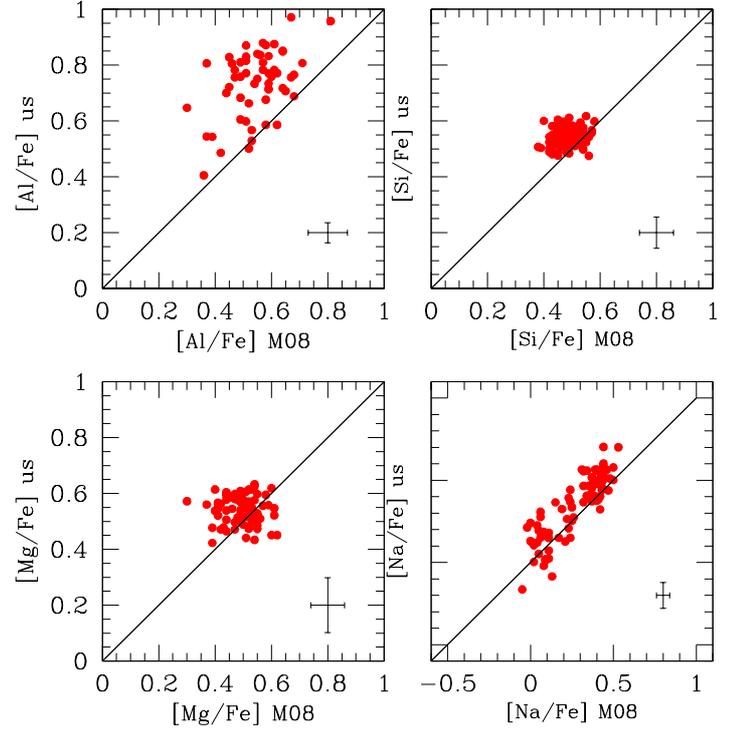}
\caption{Comparison of abundance ratios of Na, Al, Mg and Si for stars in our
sample in common with the study by Marino et al. (2008). The solid line is
the line of equality.}
\label{f:comuni61ok}
\end{figure}

We found in our sample 85 stars in common with the analysis of M~4 by Marino et
al. (2008); the comparison between our abundances of Na, Al, Mg and Si is
summarized in Fig.~\ref{f:comuni61ok}.

Taking into account differences in the scale of atmospheric parameters (based on
photometry in our case; eventually adjusted using spectroscopic parameters, in
the analysis made by M08), abundance indicators (M08 used the classical optical
doublet of Al~{\sc i} at 6696-98~\AA\ instead of the stronger doublet at
8772-73~\AA\ as in the present work), atomic parameters and the different
resolution of the spectra (higher resolution UVES spectra for M08), the
agreement is satisfactory.

Concerning 47~Tuc, recently Gratton et al. (2012b) studied a sample of 110 stars
on the red horizontal branch (HB) in 47~Tuc. The agreement in O and Na
abundances is very good: the Na-O anti-correlation as observed along the RGB and
the HB is almost indistinguishable (see Figure 5 in Gratton et al.), despite the
different abundance indicators adopted in the two studies. However, for N we
found that an offset exists, the bulk of [N/Fe] values being found between
$\sim 1.2$ and $\sim 2$ dex for HB stars in 47~Tuc.
Finally, some offset is also seen regarding Al abundances (Figure 14 in
Gratton et al.), which are derived from different doublets (7835-36~\AA\ for HB
stars and 8772-83~\AA\ here).

\subsection{Cluster analysis}

The large dataset (both in number of elements and in number of stars) 
available to us allows to perform a cluster analysis to find out whether
stars observed along the relations among light elements do form discrete groups.
As in Carretta et al. (2012a) this was done using the $k-$means algorithm
(Steinhaus 1956; MacQueen 1967) as implemented in the $R$ statistical package (R
Development Core Team 2011; http://www.R-project.org). For this exercise we
selected the abundance ratios [Al/Fe], [O/Fe], [Na/Fe], [Mg/Fe], and [N/Fe],
considering only stars in 47~Tuc and M~4 with all these elements simultaneosly
determined: 79 giants in 47~Tuc and 67 in M~4. The parameters were weighted
according to their internal errors, which are largest for O abundances, usually
derived by only one line (the associated internal errors in O are for example
about twice those relative to Na, see the Appendix in Carretta et al. 2009a).

\begin{table*}
\centering
\caption[]{Average values of abundance ratios of the different groups resulting
from the cluster analysis in 47 Tuc and M 4}
\begin{tabular}{lcccccc}
47 Tuc: 2 groups&  &        &                &                &                &         \\
\hline
group  & nr &       [N/Fe]     &        [O/Fe]    &      [Na/Fe]     &       [Mg/Fe]    &    [Al/Fe]       \\
\hline            
\hline        
1      & 41 & $+0.98 \pm 0.02$ & $+0.26 \pm 0.01$ & $+0.36 \pm 0.02$ & $+0.52 \pm 0.01$ & $+0.45 \pm 0.01$ \\
2      & 38 & $+1.11 \pm 0.01$ & $+0.00 \pm 0.02$ & $+0.59 \pm 0.02$ & $+0.52 \pm 0.01$ & $+0.64 \pm 0.01$ \\
\hline
47 Tuc: 3 groups&  &        &                &                &                &         \\
\hline
1      & 19 & $+0.87 \pm 0.01$ & $+0.31 \pm 0.02$ & $+0.28 \pm 0.02$ & $+0.51 \pm 0.02$ & $+0.40 \pm 0.01$ \\
2      & 22 & $+1.06 \pm 0.02$ & $+0.22 \pm 0.02$ & $+0.42 \pm 0.01$ & $+0.52 \pm 0.02$ & $+0.49 \pm 0.01$ \\
3      & 38 & $+1.11 \pm 0.02$ & $+0.00 \pm 0.01$ & $+0.59 \pm 0.02$ & $+0.52 \pm 0.01$ & $+0.64 \pm 0.01$ \\
       &    &                  &                  &                  &                  &                  \\
       &    &                  &                  &                  &                  &                  \\
M 4: 2 groups&  &              &                  &                  &                  &                  \\
\hline
group  & nr &       [N/Fe]     &        [O/Fe]    &      [Na/Fe]     &       [Mg/Fe]    &    [Al/Fe]       \\
\hline            
\hline        
1      & 13 & $+0.88 \pm 0.01$ & $+0.35 \pm 0.02$ & $+0.10 \pm 0.02$ & $+0.55 \pm 0.01$ & $+0.58 \pm 0.03$ \\
2      & 54 & $+1.12 \pm 0.03$ & $+0.19 \pm 0.03$ & $+0.45 \pm 0.03$ & $+0.54 \pm 0.01$ & $+0.77 \pm 0.02$ \\
\hline
M 4: 3 groups&  &        &                &                &                &         \\
\hline
1      & 13 & $+0.88 \pm 0.01$ & $+0.35 \pm 0.02$ & $+0.10 \pm 0.02$ & $+0.55 \pm 0.01$ & $+0.58 \pm 0.03$ \\
2      & 24 & $+1.21 \pm 0.01$ & $+0.26 \pm 0.02$ & $+0.42 \pm 0.02$ & $+0.53 \pm 0.01$ & $+0.73 \pm 0.03$ \\
3      & 30 & $+1.05 \pm 0.02$ & $+0.13 \pm 0.03$ & $+0.47 \pm 0.03$ & $+0.55 \pm 0.01$ & $+0.80 \pm 0.02$ \\
\hline
\end{tabular}
\label{t:clustertab}
\end{table*}

The number of subgroups to be found is a parameter that must be provided as
input to the cluster analysis when using the $k-$means algorithm. The simplest
hypothesis we could make is that two groups, i.e. stellar generations exist in
the cluster. The first one has a primordial, typical supernovae-enriched
abundance pattern; the second has a chemical composition modified by ejecta of a
fraction of stars of the first generation (the polluters).

\begin{figure}
\centering
\includegraphics[bb=100 200 428 706, clip, scale=0.52]{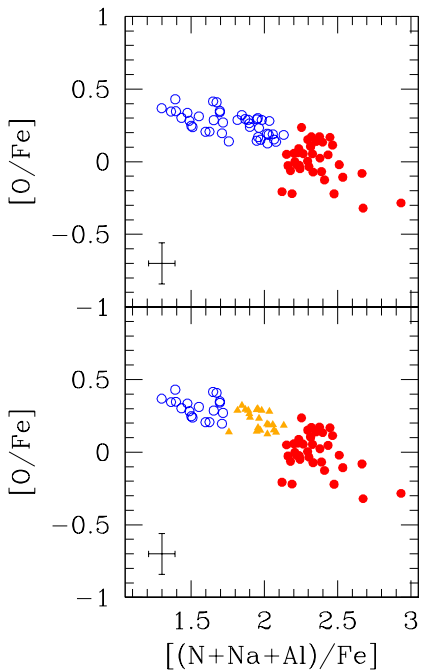}
\caption{Upper panel: results of the cluster analysis on stars in 47~Tuc with
a full set of abundances of N, O, Na, Mg, Al. The input number of groups to
be found is two. Lower panel: the same, but using three as the number of 
groups in input.}
\label{f:cluster10}
\end{figure}

The results are shown in the upper panel of Fig.~\ref{f:cluster10} for stars in
47~Tuc; we used as x-axis a parameter including all the elements that are
enhanced in the proton-capture reactions. This parameter quantifies how much the second
generation is extreme. In ordinate we display
the ratio [O/Fe] since it is the element most susceptible to depletion through
these reactions, while Mg variations are, for example, usually very tiny.
The average abundances for each group are listed in Tab.~\ref{t:clustertab}.
The algorithm basically finds in 47~Tuc two groups that corresponds to the two
generations, assigning about half of stars to each group.
A fraction of 50\% of stars of the first generation is somewhat higher than the
fraction of $\sim 30\%$ according to the definition of primordial P fraction
given in Carretta et al. (2009a).

\begin{figure}
\centering
\includegraphics[bb=100 200 428 706, clip, scale=0.52]{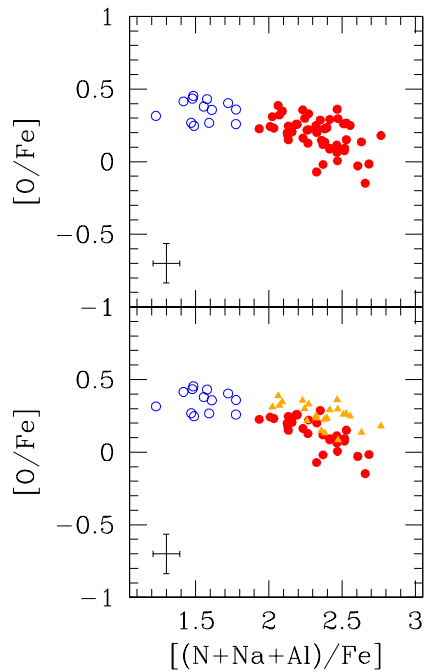}
\caption{The same as in Fig.~\ref{f:cluster10} for M~4.}
\label{f:cluster61}
\end{figure}

However, looking at the upper panel in Fig.~\ref{f:cluster10} we note that
beside the rather clear separation between stars in the first (open circles)
and in the second groups (filled circles) there is some evidence for a
discontinuity in the distribution also $within$ the first group, located
approximatively at [(N+Na+Al)/Fe]=1.8 dex.
To verify how sound is this feature, we ran again the cluster analysis,
forcing the algorithm to find three groups.

The results are presented in the lower panel of Fig.~\ref{f:cluster10}: the stars of the
previous first generation are now split into two groups and the putative
intermediate population detected above nicely falls into a unique group (filled
triangles), apart from a single outlier. With this separation, the fraction of
stars with primordial composition is now 24\%, while the intermediate and
extreme populations summed together reach 76\% of the sample, reassuringly close
to the typical fractions derived in Carretta et al. (2009a) using only O and Na
abundances.
This is not surprising, since the positioning of the GIRAFFE fibers with
the HR21 grating in the present study, was done using the same configuration
used for Na (HR11) by Carretta et al. (2009a). As explained in that paper, these
configurations, due to the limitations of the Oz-Poz positioner of FLAMES, are
not the best suited to study the radial dependence of the fraction of stars in
the P, I and E components. This also makes difficult a direct comparison with
methods based on photometric parameters, like those in Milone et al. (2012),
more able to study radial distributions, but with different sensitivities on
chemistry and age. For instance, we note that in Milone et al., three
populations were found among subgiants, but not among RGB stars. Also, we are not 
sure that the three populations found along the RGB in the present study by means
of chemical composition are the same found along the SGB. 

For M~4 the case is much simpler. The cluster analysis with only two groups
as input separates quite well the first and second generation stars, as
evident in the upper panel of Fig.~\ref{f:cluster61} where the gap along the
"secondness" parameter is very clear.
Hence, it is not surprising that the cluster analysis fails to find three 
distinct subgroups (lower panel in
Fig.~\ref{f:cluster61}): in this case an "intermediate" group is found by the
algorithm, but it is barely separated from the bulk of second generation stars.
A look at the abundances listed in Tab.~\ref{t:clustertab} shows that this
putative intermediate population has N abundances too high, with respect to
the extreme groups, and the likely explanation is that these are the warmest
stars, responsible for the "flaring" of the scatter in N seen above.
This is simply an effect of the analysis and not a true feature in this
cluster.

\section{Discussion and conclusions}

The present study highlights the advantage of having data for a large set of
elements involved in the network of proton-capture reactions of H-burning at
high temperature. It is very useful to have a simultaneous look at all
relations among light elements because the depletions observed for O and Mg, as
well as the enhancements detected in N, Na, and Al allow probing the
stratigraphy within the H-burning region, reaching more and more hot layers.
In turn, this translates into evidence of more massive stars being
responsible as the source of intra-cluster pollution.

This large harvest of data can then be exploited to understand
whether the formation of globular clusters was characterized by a single
second episode or multiple episodes of star formation. Of course, when the
observational evidence points toward only two distinct groups, as found in M~4
in the present and previous analyses (Ivans et al. 1999, Marino et al. 2008,
Carretta et al. 2009a), the question must necessarily
remain unanswered and the scenario is undetermined.

More interesting is the case when more than two groups are found. Leaving aside
the peculiar case of $\omega$ Cen, the most noticeable case among normal GCs is
NGC~2808. We know that this cluster shows three distinct main sequences
(D'Antona et al. 2005; Piotto et al. 2007), has an unusual HB, bimodal and with
a long blue tail (Walker 1999) that has been connected to different He
enrichments (D'Antona and Caloi 2004), and shows a very extended Na-O
anti-correlation, with some evidence of multiple "blobs" (Carretta et al. 2006).

Another more recent case is NGC~6752, where with our survey of Al abundances we
found the evidence of three distinct stellar populations with different chemical
composition. Moreover, from the correlation between depleted species such as O
and Mg (clear but not linear) we were able to show that the composition of the
intermediate population cannot be obtained by simply mixing primordial matter
with material having the chemical composition of the extreme population.

The same approach can be followed to try to better understand our present
finding in 47~Tuc. Using a simple dilution model (see Carretta et al. 2009a,
Gratton et al. 2010) we searched for the dilution factors able to reproduce the
intermediate group by mixing the typical composition of stars from the other two
groups. If the model is correct we should obtain the same factor, regardless of
our choice of the element to be used. We derived that dilution factors 0.71, 
0.64, and 0.63 (basically, the same factor, within the uncertainties) are 
required for Al, O and Na, respectively. The Mg abundance variations are so 
small that virtually any factor applies. Our nominal N abundances, which
were obtained assuming a constant value of [C/Fe]=-0.6 for all stars, require a
dilution factor of 0.29 that is quite different from the values obtained for
the other elements. However, this result depends on the adoption of a constant [C/Fe]
and a different conclusion is obtained by assuming that the C abundance is different 
in the three groups. This is not at all an ad hoc assumption, since several studies 
showed that the [C/Fe] ratio is higher by $\sim 0.3$~dex in CN-weak (Na-poor, O-rich) 
stars than in CN-strong (Na-rich, O-rich) stars (see Norris \& Freeman 1982, 
Norris \& Cottrell 1979; Smith et al. 1989; Brown et al. 1990; Briley 1997; 
Carretta et al. 2005). Were the C abundance in group 2 be also intermediate
between those of groups 1 and 2, it would be possible to reproduce the strength
of the CN lines with a N abundance compatible with a pollution scenario.
Direct determination of C abundances for the stars in our sample would solve
this ambiguity, and clarify if dilution may explain the pattern seen in 47~Tuc.
The same observation could also be used to find if the sum of C+N+O abundances
is constant or not in 47 Tuc. We only note here that if we combine the C abundances
for CN-weak and CN-strong RGB stars by Brown et al. (1990) and Briley (1997) 
([C/Fe]$\sim -0.1$ and [C/Fe]$\sim -0.3$, respectively) with our CN and O
abundances, the sum of C+N+O abundances would be the same for the two extreme
groups within 0.1~dex. This result agrees quite well with the study of the
C+N+O sum in unevolved cluster stars in NGC~6752, NGC~6397, and 47~Tuc made by
Carretta et al. (2005). 

Another problem in 47 Tuc is to find the progeny of the subgiant (SGB) stars C+N+O enhanced,
the so called ``SGII", hypothesised by Di Criscienzo et al. (2010)  that are 10\% of the stars.
More investigations are needed to have a self-consistent picture on this cluster.

A promising approach could be to look for the full set of abundances, from C, N,
O to the light and heavy, neutron-capture elements, especially if measured in
the same stars. However this is generally not available at the moment, so no
stringent conclusions can be drawn. For instance, Worley et al. (2010) present a
list of studies regarding heavy elements in 47~Tuc. Only in a few of them, and
only for a limited number of stars, abundances for the whole set of elements is
available (e.g, 12 TO and SGB stars in James et al. 2004, when combined with
Carretta et al. 2005; four bright giants in Brown \& Wallerstein 1992). An
interesting approach based on medium resolution spectra of a large number of
targets is presented in Worley \& Cottrel (2012), who obtained a measure of the
CN excess for about 100 giants, with almost no overlap with our present sample.
However, they did not measure O abundances and published results of the
abundance analysis of other elements  only for a small subsample of 13 bright
stars. In all the above-mentioned studies there seems to be an homogeneous
pattern among elements produced by neutron-capture processes, and this
conclusion seems to be a severe drawback for the hypothesis of a significant
contribution of  polluted matter by low-mass  AGB stars in this cluster. We
stress however that all the existing samples are too small to permit a
separation of groups as we do in the present paper using the cluster analysis
technique. Observational efforts should be devoted to acquiring the most
complete set of key elements abundances in significant samples of stars.

Our results show that 47~Tuc is the third normal globular cluster  
after NGC~2808 (D'Antona et al. 2005, Piotto et al. 2007, Carretta et al.
2006) and  NGC~6752 (Yong et al. 2008, Carretta et al. 2012a), where at least
three 
groups of stars with homogeneous chemical composition (within each
group) can be separated. The implication seems
to be that for a fraction of GCs the scenario of cluster formation 
must include multiple bursts where second generation stars are produced.
In each, the interplay of matter processed by
polluters of different (obviously decreasing with time) average masses, and
pristine material does combine to give the observed chemical pattern in each
group. By studying the detailed abundance distribution of proton-capture
elements we can then provide important constraints to the theoretical models of
cluster formation.

All these ingredients must be carefully considered and tuned into existing and
future theoretical models. Homogeneous analysis of several proton-capture
elements in large sample of stars such as the one presented in this study may
provide a fruitful way to contribute useful constraints to these models.

\begin{acknowledgements}
This work was partially funded by the PRIN INAF 2009 grant CRA 1.06.12.10
(``Formation and early evolution of massive star clusters", PI. R. Gratton),
and by the PRIN INAF 2011 grant ``Multiple populations in globular clusters:
their role in the Galaxy assembly" (PI E. Carretta). We thank
\v{S}ar\={u}nas Mikolaitis for sharing with us the line list for CN provided
by B. Plez.
This research has made use of the SIMBAD database (in particular  Vizier),
operated at CDS, Strasbourg, France and of NASA's Astrophysical Data System.
\end{acknowledgements}


\begin{thebibliography}{}

\bibitem[]{} Arnould, M., Goriely, S., Jorissen, A. 1999, A\&A, 347, 572 
\bibitem[]{} Bragaglia, A., Carretta, E., Gratton, R.G. et al. 2010, ApJ, 720,
  L41 
\bibitem[]{} Briley, M.M. 1997, AJ, 114, 1051
\bibitem[]{} Brown, J.~A., Wallerstein, G. 1992, AJ, 104, 1818
\bibitem[]{} Brown, J.A., Wallerstein, G., Oke, J.B. 1990, AJ, 100, 1561
\bibitem[]{} Carlberg, J.K., Cunha, K., Smith, V.V., Majewski, S.R. 2012, ApJ, 757, 109
\bibitem[]{} Carretta, E., Gratton, R.G., Lucatello, S., Bragaglia, A., Bonifacio, P. 2005,
A\&A, 433, 597 
\bibitem[]{} Carretta, E., Bragaglia, A., Gratton, R.G., D'Orazi, V., Lucatello,
 S. 2009c, A\&A, 508, 695 
\bibitem[]{} Carretta, E., Bragaglia, A., Gratton R.G., Leone, F.,
Recio-Blanco, A., Lucatello, S. 2006, A\&A, 450, 523 
\bibitem[]{} Carretta, E., Bragaglia, A., Gratton, R.G., \& Lucatello, S.
  2009b,  A\&A, 505, 139 (Paper VIII)  
\bibitem[]{} Carretta, E., Bragaglia, A., Gratton, R.G., Lucatello, S., D'Orazi,
  V. 2012a, ApJ, 750, L14 
\bibitem[]{} Carretta, E., D'Orazi, V., Gratton, R.G., Lucatello, S. 2012b,
  A\&A, 543, A117 
\bibitem[]{}  Carretta, E., Bragaglia, A., Gratton, R.G., Recio-Blanco, A.,
  Lucatello, S., D'Orazi, V., \& Cassisi, S. 2010a, A\&A, 516, 55 
\bibitem[]{} Carretta, E., Lucatello, S., Gratton, R.G., Bragaglia, A., D'Orazi,
  V. 2011, A\&A, 533, 69 
\bibitem[]{} Carretta, E., Bragaglia, A., Gratton, R.G. et al. 2009a, A\&A,
  505, 117  
\bibitem[]{} Carretta, E., Gratton R.G., Bragaglia, A., Bonifacio, P. \& 
  Pasquini, L. 2004, A\&A, 416, 925 
\bibitem[]{} Carretta, E., Gratton R.G., Lucatello, S., Bragaglia, A., 
  Bonifacio, P. 2005, A\&A, 433, 597

\bibitem[]{} D'Antona, F., Caloi, V. 2004, ApJ, 611, 871 
\bibitem[]{} D'Antona, F., Bellazzini, M., Caloi, V., Fusi Pecci, F., Galleti,
 S., Rood, R.T. 2005, ApJ, 631, 868
\bibitem[]{} D'Ercole, A., D'Antona, F., Vesperini, E. 2011, MNRAS, 415, 1304
\bibitem[]{} Decressin, T., Meynet, G., Charbonnel C. Prantzos, N.,\&  Ekstrom,
  S. 2007, A\&A, 464, 1029 
\bibitem[]{} de Mink, S.E., Pols, O.R., Langer, N., Izzard, R.G. 2009, A\&A,
  507, L1 
\bibitem[]{} Denisenkov, P.A.,\&  Denisenkova, S.N. 1989, A.Tsir., 1538, 11
\bibitem[]{} Di Criscienzo, M., Ventura, P., D'Antona, F., Milone, A., Piotto,
G. 2010, MNRAS, 408, 999 
\bibitem[]{} D'Orazi, V., Gratton, R.G., Pancino, E., et al. 2011, A\&A 534, A29
\bibitem[]{} Gratton, R.G. 1988, Rome Obs. Preprint Ser., 29 
\bibitem[]{} Gratton, R.G., Carretta, E. \& Bragaglia, A. 2012a, A\&A Rev., arXiv:1201.6526
\bibitem[]{} Gratton, R.G., Lucatello, S., Sollima, A. et al. 2012b, submitted to A\&A
\bibitem[]{} Gratton, R.G., Carretta, E., Bragaglia, A., Lucatello, S., D'Orazi,
  V. 2010, A\&A, 517, 81
\bibitem[]{} Gratton, R.G., Sneden, C., \& Carretta, E. 2004, ARA\&A, 42, 385
\bibitem[]{} Gratton, R.G., Bonifacio, P., Bragaglia, A. et al. 2001, A\&A, 369, 87
\bibitem[]{} Harris, W., 1996, AJ, 112, 1487
\bibitem[]{} Hinkle, K., Wallace, L., Valenti, J., Harmer, D. 2000, Visible and Near Infrared Atlas 
of the Arcturus Spectrum 3727-9300 \AA (San Francisco: ASP)
\bibitem[]{} Ivans, I.I., Sneden, C., Kraft, R.P., Suntzeff, N.B, Smith, V.V.,
  Langer, G.E., Fulbright, J.P. 1999, AJ, 118, 1273 
\bibitem[]{} James, G., Fran{\c c}ois, P., Bonifacio, P., Carretta, E., Gratton,
 R.G., Spite, F. 2004, A\&A, 427, 825 
\bibitem[]{} Kraft, R.~P. 1994, PASP, 106, 553 
\bibitem[]{} Kurucz, R.L. 1993, CD-ROM 13, Smithsonian Astrophysical Observatory, Cambridge
\bibitem[]{} Langer, G.E., Hoffman, R., \& Sneden, C. 1993, PASP, 105, 301
\bibitem[]{} MacQueen, J. B. 1967, Mathematical Statistics and Probability, University of California Press., 281.  
\bibitem[]{} Marino, A.F., Villanova, S., Piotto, G., Milone, A.P., Momany, Y.,
  Bedin, L.R., Medling, A.M. 2008, A\&A, 490, 625 
\bibitem[]{} Martell, S.L. 2011, AN, 332, 467 
\bibitem[]{} Milone, A., Piotto, G., Bedin, L. et al. 2012, ApJ, 744, 58 
\bibitem[]{} Norris, J.E., Cottrell, P.L. 1979, ApJ, 229, 69 

\bibitem[]{} Norris, J., Freeman, K.C. 1979, ApJ, 230, L179 
\bibitem[]{} Norris, J.E., Freeman, K.C. 1982, ApJ, 254, 143 

\bibitem[]{} Piotto, G. et al. 2007, ApJ, 661, L53 
\bibitem[]{} Prantzos, N., Charbonnel, C. 2006, A\&A, 458, 135 
\bibitem[]{} R Development Core Team, 2011, R: A language and environment for
statistical computing, Vienna, Austria, ISBN 3-900051-07-0
\bibitem[]{} Ramirez, S. \& Cohen, J.G. 2002, AJ, 123, 3277
\bibitem[]{} Smith, G.H., Bell, R.A., Hesser, J.E. 1989, ApJ, 341, 190 

\bibitem[]{} Smith, G.H., Briley, M.M. 2005, PASP, 117, 895 
\bibitem[]{} Steinhaus, H. 1956, Bull. Acad. Polon. Sci. 4, 801
\bibitem[]{} Ventura, P. D'Antona, F., Mazzitelli, I., \& Gratton, R. 2001, ApJ,
  550, L65 
\bibitem[]{} Walker, A.R. 1999, AJ, 118, 432 
\bibitem[]{} Worley, C.~C.,Cottrell, P.~L. 2012, PASA, 29, 29 
\bibitem[]{} Worley, C.~C., Cottrell, P.~L., McDonald, I., van Loon, J.~T. 
 2010, MNRAS, 402, 2060
\bibitem[]{} Yong, D., Grundahl, F.,  Nissen, P.E., Jensen, H.R., Lambert, D.L. 2005,
 A\&A, 438, 875
\bibitem[]{} Yong, D., Grundahl, F., Johnson, J.A., Asplund, M. 2008, ApJ, 684,
  1159 

\end{thebibliography}
\end{document}